\documentclass[3p]{elsarticle}
\usepackage[T1]{fontenc}
\usepackage[utf8]{inputenc}
\usepackage{lmodern}
\usepackage{epsfig} 
\usepackage{graphicx}
\usepackage{amsmath}
\usepackage{color}
\usepackage{slashed}
\usepackage{subfig}
\usepackage{mathtools}
\usepackage{natbib}
\usepackage{lineno}
\usepackage{setspace}

\usepackage{graphicx}
\usepackage{dcolumn}
\usepackage{bm}
\usepackage{float}
\usepackage[polish,USenglish]{babel}
\biboptions{sort&compress}

\begin{document}

\begin{frontmatter}

\title{The J-PET detector - a tool for precision studies of
ortho-positronium decays}
\author[1,2]{K. Dulski}
\address[1]{Faculty of Physics, Astronomy and Applied Computer 
Science, Jagiellonian University, 
30-348 Cracow, Poland}
\address[2]{Total-Body Jagiellonian-PET Laboratory, Jagiellonian 
University, Poland}
\author[1,3]{S.~D. Bass}
\address[3]{Kitzb\"uhel Centre for Physics,
Kitzb\"uhel, Austria}
\author[1,2]{J. Chhokar}
\author[1,2]{N. Chug}
\author[4]{C. Curceanu}
\address[4]{INFN, Laboratori Nazionali di Frascati, 00044 Frascati, Italy}
\author[1,2]{E.~Czerwiński}
\author[1,2]{M.~Dagdar}
\author[5]{J.~Gajewski}
\address[5]{Institute of Nuclear Physics Polish Academy of Sciences, 31-342 Cracow, Poland}
\author[1,2]{A.~Gajos}
\author[6]{M.~Gorgol}
\address[6]{Institute of Physics, Maria Curie-Sk\l{}odowska University, 20-031 Lublin, Poland}
\author[4]{R.~Del~Grande}
\author[7]{B.~C.~Hiesmayr}
\address[7]{Faculty of Physics, University of Vienna, 1090 Vienna, Austria}
\author[6]{B.~Jasińska}
\author[1]{K.~Kacprzak}
\author[1,2]{Ł.~Kapłon}
\author[1,2]{H.~Karimi}
\author[1]{D.~Kisielewska}
\author[8]{K.~Klimaszewski}
\address[8]{Department of Complex Systems, National Centre for Nuclear Research, 05-400 Otwock-\'Swierk, Poland}
\author[8]{P.~Kopka}
\author[1]{G.~Korcyl}
\author[8]{P.~Kowalski}
\author[1]{T.~Kozik}
\author[1,2]{N.~Krawczyk}
\author[9]{W.~Krzemień}
\address[9]{High Energy Physics Division, National Centre for Nuclear Research, 05-400 Otwock-\'Swierk, Poland} 
\author[1,2]{E.~Kubicz}
\author[10]{P.~Małczak}
\address[10]{2nd Department of General Surgery, Jagiellonian University Medical College, Cracow, Poland}
\author[1,11]{M.~Mohammed}
\address[11]{Department of Physics, College of Education for Pure Sciences, University of Mosul, Mosul, Iraq}
\author[1,2]{Sz.~Niedźwiecki}
\author[1]{M.~Pałka}
\author[1,2]{M.~Pawlik-Niedźwiecka}
\author[10]{M.~Pędziwiatr}
\author[8]{L.~Raczyński}
\author[1,2]{J.~Raj}
\author[5]{A.~Ruci\'nski}
\author[1,2]{S.~Sharma}
\author[1,2]{Shivani}
\author[8]{R.Y.~Shopa}
\author[1,2]{M.~Silarski}
\author[1]{M.~Skurzok}
\author[1,2]{E.Ł.~Stępień}
\author[1,2]{F.~Tayefi}
\author[8]{W.~Wiślicki}
\author[6]{B.~Zgardzińska}
\author[1,2]{P.~Moskal}

\doublespacing
	
\begin{abstract}
The J-PET tomograph is constructed from plastic scintillator strips arranged axially in concentric cylindrical layers. 
It enables investigations of positronium decays by measurement 
of the time, position, polarization and energy deposited by photons in the scintillators, 
in contrast to studies conducted so far with crystal and semiconductor based detection systems where the key selection of events is based on the measurement of the photons' energies. 
In this article we show 
that the J-PET tomograph system 
is capable of exclusive 
measurements of the decays of ortho-positronium atoms.
We present the first positronium production results, 
its lifetime distribution measurements and discuss
estimation of the influence of various background sources. 
The tomograph's performance demonstrated here makes it 
suitable for 
precision studies of positronium decays including entanglement
of the final state photons,
positron annihilation lifetime spectroscopy 
plus 
molecular imaging diagnostics.
\end{abstract}


 \end{frontmatter}

\section{Introduction}
\doublespacing

The J-PET detector,
Jagiellonian Positron Emission Tomograph,
developed 
in Cracow
\cite{Moskal:2014sra,Moskal:2015nim}
is a new tool for precision studies of positronium decays.
The J-PET programme
focusses on
fundamental studies of positronium decays~\cite{X5,X4,X2,Yamazaki:2009hp,Y1,Y2,X1,Bernreuther:1988tt}, 
on positron annihilation lifetime spectroscopy (PALS)~\cite{Moskal:2018wfc} 
as well as medical diagnostics~\cite{Mos_PETClinics,VanMosKar_EJNMMI,SloPanGer_SNM,Badawi_JNM,Karp_JNM}.
The J-PET detector is a new PET device based on plastic scintillators
designed for total body scanning in medicine
as well as for
biological applications~\cite{Moskal:2016a,Niedzwiecki:2017app,Kowalski:2018pmb,Moskal:2018wfc} 
and fundamental physics research \cite{Moskal:2016moj} 
with detection of positronium 
via Compton rescattered photons in the detector.

In this paper we present the first ortho-positronium, o-Ps, production results 
from J-PET 
as well as lifetime distributions that can be compared with 
theoretical predictions 
coming 
from previous simulations~\cite{Kaminska:2016fsn,Moskal:2018pus}.
These results, especially the identification of o-Ps decays, are
the first obtained using a~detector based solely on plastic scintillators. 
Previous studies of o-Ps decays were conducted
with crystal and semiconductor based detection systems. 
In addition, by using specific and realistic detector simulations we are able to estimate the extent to which we can reduce the background influence on the distributions.

Positronium, an ``atom'' consisting of an electron and a positron,
is described by bound state QED 
with very small radiative corrections from Quantum Chromodynamics
and weak interaction effects in the Standard Model.
Positronium comes in two ground states,
$^1 S_0$ para-positronium, denoted p-Ps, 
with spin equal to zero
and 
$^3 S_1$ ortho-positronium, 
denoted o-Ps, 
with spin equal to one.
\hbox{p-Ps} is slightly lighter by 0.84 meV due to 
the interaction between the electron and positron spins 
and also the existence of the o-Ps one-photon virtual annihilation processes.
Spin-zero p-Ps has a mean
lifetime of 125 picoseconds 
and spin-one o-Ps has a mean lifetime of 142 nanoseconds
in vacuum.
%

%
%


Reviews of positronium physics are given 
in~\cite{Cassidy:2018tgq,Bass:2019ibo}.
Measurements of positronium decay rates are so far consistent 
with bound state QED theoretical predictions
performed in the framework of non-relativistic QED \cite{Adkins:2002fg}.
However, the present experimental uncertainties 
$\sim {\cal O}(10^{-4})$
are much larger than the theoretical uncertainties
(assuming the non-relativistic QED approach)
by a factor of 100 for o-Ps and by 10,000 for p-Ps,
thus calling for increased experimental precision.
%
Additional observables \cite{Moskal:2016moj} in positronium decays involving correlations between the positronium's spin and the momenta and polarization of the photons emitted in the 
decays can be used to test discrete symmetries $C$, $CP$ and $CPT$ in o-Ps bound state decays. 
Information about the photons' polarization is deduced
from the observation of the Compton scatterings between 
pair of scintillators in the detector \cite{Moskal:2018pus, Nikodem:2019}.
Charge conjugation invariance in this system has been tested
up to the level of $10^{-6}$ \cite{X5,X4,X2} and
the symmetries $CP$ \cite{Yamazaki:2009hp,Y1}
and $CPT$ \cite{Y2,X1}
have so far been tested up to ${\cal O}(10^{-3})$.

Possible invisible decays of positronium are also an 
interesting topic of investigation with the J-PET 
detector~\cite{Krzemien_Elena_Kacprzak_2020_Acta}.
Mirror matter models of dark matter allow a branching 
ratio for the invisible decay of o-Ps in vacuum 
to mirror particles
up to about $2 \times 10^{-7}$, 
below the present experimental bound of $3.0 \times 10^{-5}$
\cite{Vigo:2018xzc,Vigo:2020}.
With J-PET, the experimental accuracy on the total decay rate
and discrete symmetries observables is expected to be 
improvable by an order of magnitude in precision~\cite{Kaminska:2016fsn,Moskal:2018pus,Krzemien_Elena_Kacprzak_2020_Acta}.

Studies of positronium decays have important applications also
in medicine and biology.
Inside biological materials the positronium mean lifetime and 
formation probability depend significantly
on the healthiness of the tissue material, 
its nanostructure and concentration of bioactive molecules,
in effect allowing us to
distinguish between healthy
and altered tissue \cite{nrp:2019, bura:2020app, jasin:2017appa}. 
These factors are indicative of the stage of development of metabolic 
disorders of human tissues. 
Thus, positronium decay studies can provide new input in medical diagnosis
\cite{nrp:2019,Moskal:2018wfc,EJNMMI_arXiv:1911.06841,Hiesmayr_ScientiReport2017,Hiesmayr_ScientificReport2019}.
Other important applications are 
to chemistry and material characterization \cite{Mat1, Mat2}.

The plan of the paper is as follows.
In Section 2 we describe the J-PET detector setup.
Section 3 discusses the event selection method and results.
Conclusions are given in Section 4.

\section{J-PET detector and positronium production method}

\begin{figure}[h!]
\centering
\includegraphics[width=0.335\textwidth]{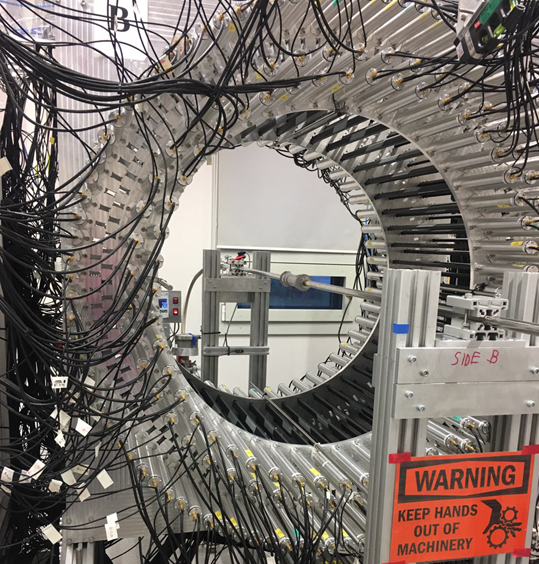}
\includegraphics[width=0.335\textwidth]{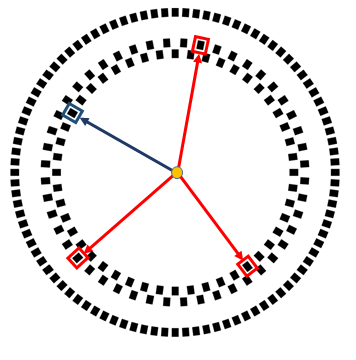}
\caption{\label{fig:JPET} (left) Photograph of the J-PET detector with the annihilation chamber placed in the centre of the detector with a pipe installed on the detector axis. (right) Schematic cross section of the J-PET detector. Black rectangles indicate the cross section of scintillator strips arranged in three concentric cylindrical layers. 
The arrows indicate annihilation photons (red) from ortho-positronium and the de-excitation photon (blue) emitted by the $^{22}$Na source located at the centre of the detector.}
\end{figure}

The J-PET detector (Fig.~\ref{fig:JPET}-left) is built from polyvinyl toluene base EJ-230 plastic scintillator strips~\cite{EJ230} with dimensions 50$\times$1.9$\times$0.7 cm$^3$, arranged in three concentric cylindrical layers with radius 42.5 cm, 46.75 cm and 57.5 cm~\cite{Niedzwiecki:2017app} (Fig.~\ref{fig:JPET}-right). Each scintillator is connected from both sides with vacuum photomultipliers R9800~\cite{R9800}. It allows one to reconstruct the position along the scintillator based on the time difference between signals registered at photomultipliers connected to the same scintillator~\cite{Neha_reco, Lech_reco}. The scanner is equipped with the dedicated solely digital field programmable gate array platform (FPGA)~\cite{FPGA2} and trigger-less data acquisition system based on TRB3 hardware~\cite{FPGA1}, optimised for detection of photons originating from positronium decay and photons
from the nuclear deexcitations in the energy region of 1~MeV. Signals from the photomultipliers are transferred to the acquisition system and scanned at four different thresholds - 80 mV, 160 mV, 240 mV and 320 mV.
Next, the data in the form of list-mode is further transferred to the analysis software.
Analysis is performed with a~dedicated analysis and simulation framework~\cite{Framework,Framework2}. Photons below 10 MeV 
energy interact in the plastic scintillators predominantly via the Compton effect. Therefore, the information about the 
energy of the photons is not available directly, and the whole data selection and analysis chain is based on the determination of the time and position of the photon's interaction and its energy deposition. The time and position resolutions achieved with the J-PET detector are about 380~ps and 4.6~cm (in FWHM), respectively~\cite{Niedzwiecki:2017app}. This corresponds to the angular resolution equal to about 2.4 degree (in FWHM) \cite{Kaminska:2016fsn}. Comparing to other large area detectors like Gammasphere or APEX array, the J-PET detector is characterized by better time resolution (4.6 ns for Gammasphere \cite{X2}, 2 ns for APEX array \cite{APEX}). In addition, due to trigger-less data acquisition system, the J-PET detector is capable of overcome the limitation of pileup events coming from higher rates of the positronium production, which allows to use higher source activities (10 MBq for the J-PET compared to 0.37 MBq for Gammasphere and APEX array). It provides the possibility to obtain higher statistic, which allows to obtain much more accurate results in the fundamental physics studies.

\begin{figure}[h]
\centering
\includegraphics[width=0.6\textwidth]{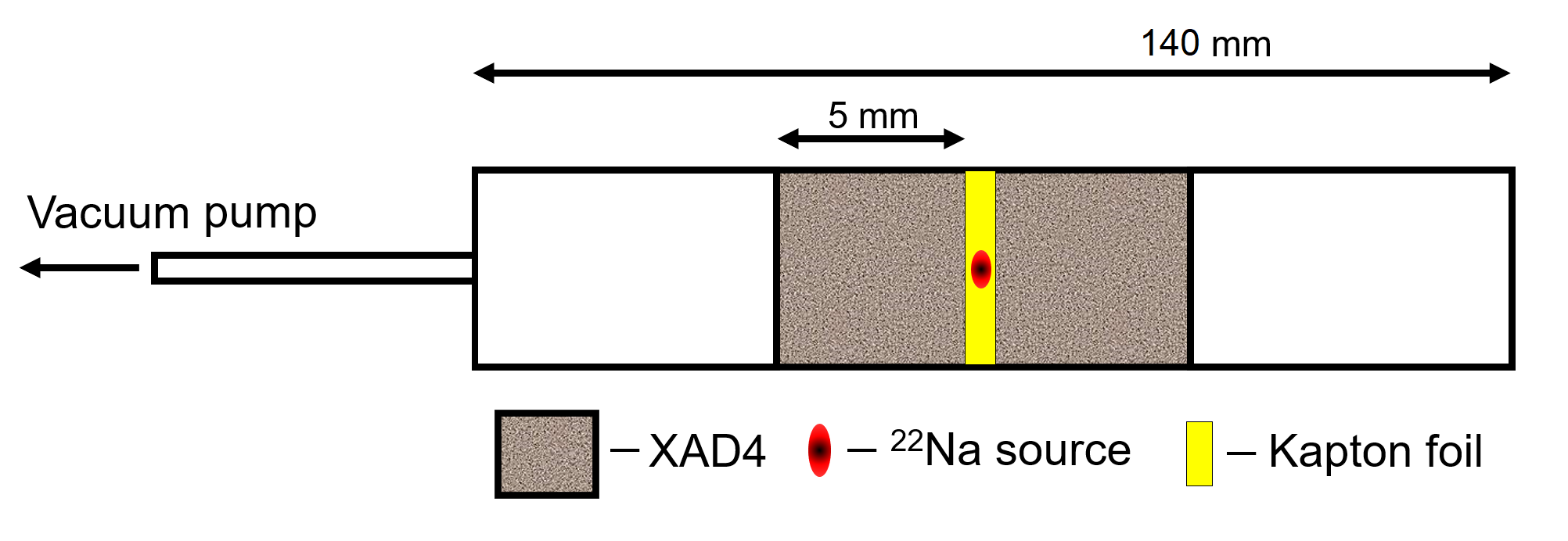}
\caption{\label{fig:AnniChamb} Scheme of  
the annihilation chamber with the $^{22}$Na radioactive source surrounded with XAD4 material. The chamber is connected on one side to the vacuum pump. Chamber details are published in \cite{Gorgol_APB}.}
\end{figure}

Following part is the description of the positronium setup used in the measurement. Positronium atoms are created from positrons (emitted from the $^{22}$Na isotope) and electrons from the porous polymer XAD4~\cite{sigma}.
The polymer was placed inside the annihilation chamber 
around 
a $^{22}$Na source with activity of 1 MBq, wrapped in a~$8~\mu$m thin Kapton foil, as indicated in Fig.~\ref{fig:AnniChamb}. The annihilation chamber was installed in the centre of the J-PET detector as can be seen in the photograph in the left panel of Fig.~\ref{fig:JPET}. 
The air was pumped out of the polymer, and during the measurement the pressure in the annihilation chamber was sustained at the level of 10~Pa. The XAD4 polymer was chosen as a positronium production medium since it is characterized by a~high fraction of positron-electron decays into three photons, $f_{3 \gamma} = 28.9\%$~\cite{Jasin}. 

The right panel of Fig.~\ref{fig:JPET} shows an example of an event in which the three photons from o-Ps annihilation, as well as the de-excitation photon from the $^{22}$Na source were registered. 
Exclusive measurements of the interaction's position (and hence relative angles) of the three annihilation photons enable one to reconstruct the full event's kinematics (momenta of all photons)~\cite{Kaminska:2016fsn}. 
The additional measurement of the de-excitation photon from the $^{22}$Na decay chain ($^{22}$Na $\to$ $^{22}$Ne$^*$ e$^+$ $\nu$ $\to$ $^{22}$Ne $\gamma$ e$^+$ $\nu$) allows one to determine the lifetime of the positronium atom. However, the finite geometrical acceptance and detection efficiency of the J-PET detector, and the multiple Compton scatterings, 
also mean that there are many other possible event types 
with a similar signature.
Therefore we need
sophisticated selection procedures that we describe in the next section.

In order to estimate the influence of various background sources that could survive selection procedures, Monte Carlo simulations were preformed for the experimental setup described above. Simulations were done with dedicated software for the J-PET detector \cite{Kaminska:2016fsn, Framework, Framework2}, and analysed using the same analysis procedure as for the experimental data. The positron lifetime distribution for the XAD4 material was assumed as described in \cite{Jasin}.

\section{Selection of ${\rm \hbox{o-Ps}}$ $\to$ 3$\gamma$ events}

\begin{table}[t]
\centering
\caption{Different types of observed events. Arrows denote registered annihilation photons (solid red), not registered annihilation photons (dashed gray), and deexcitation photon (dotted blue). }\label{Tab:TypesOfEvents}
\begin{tabular}{|c|c|c|}
\hline
\textbf{(I)} & \textbf{(II)} & \textbf{(III)}\\
\hline
& & \\
\multicolumn{3}{|c|}{Annihilation into 3 photons}\\
& & \\
\includegraphics[width=0.135\textwidth]{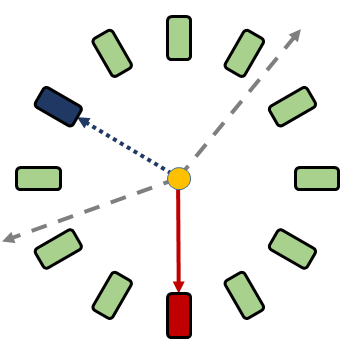} &  \includegraphics[width=0.145\textwidth]{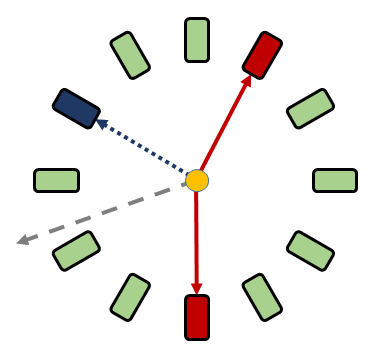} & \includegraphics[width=0.135\textwidth]{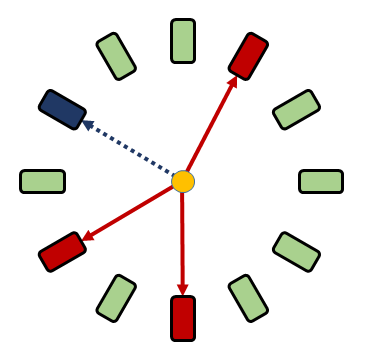} \\
\hline
& & \\
\multicolumn{3}{|c|}{Annihilation into 2 photons}\\
& & \\
\includegraphics[width=0.135\textwidth]{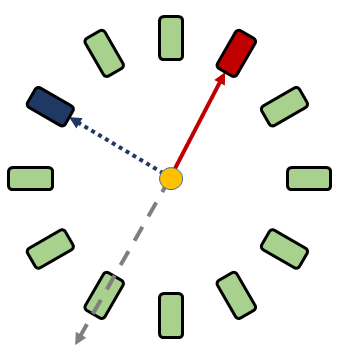} &  \includegraphics[width=0.145\textwidth]{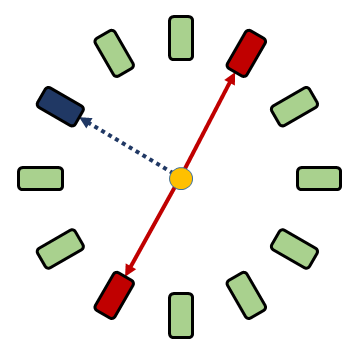} & \\
\hline
\end{tabular}
\end{table}

The positron-electron annihilation predominantly leads to the production of two or three photons. It may proceed directly (e$^+$e$^-$ $\to$ 2$\gamma$(3$\gamma)$) or via formation of a~positronium atom (e$^+$e$^-$ $\to$ Ps $\to$ 2$\gamma$(3$\gamma$)). When requiring registration of the de-excitation photon
(needed for the determination of the positron lifetime in the XAD4 material) we may define three classes of events with one (I), two (II) or three (III) registered annihilation photons, as it is graphically depicted in Table~\ref{Tab:TypesOfEvents}. In vacuum o-Ps decays only to three photons (due to C symmetry conservation). However, in the intermolecular voids of the material it may decay also into two back-to-back photons via conversion~\cite{Step_PCCP} or pick-off~\cite{Pick_off} processes.  
The inclusive measurements using categories I, II  are typically used both in PALS~\cite{Jasin,Step_PCCP,PALS_Review} and in most of the so far conducted experiments aimed at 
studies of discrete symmetries~\cite{X4,Yamazaki:2009hp,Y1,Y2}.

\begin{figure}[b!]
\centering
\includegraphics[width=0.6\textwidth]{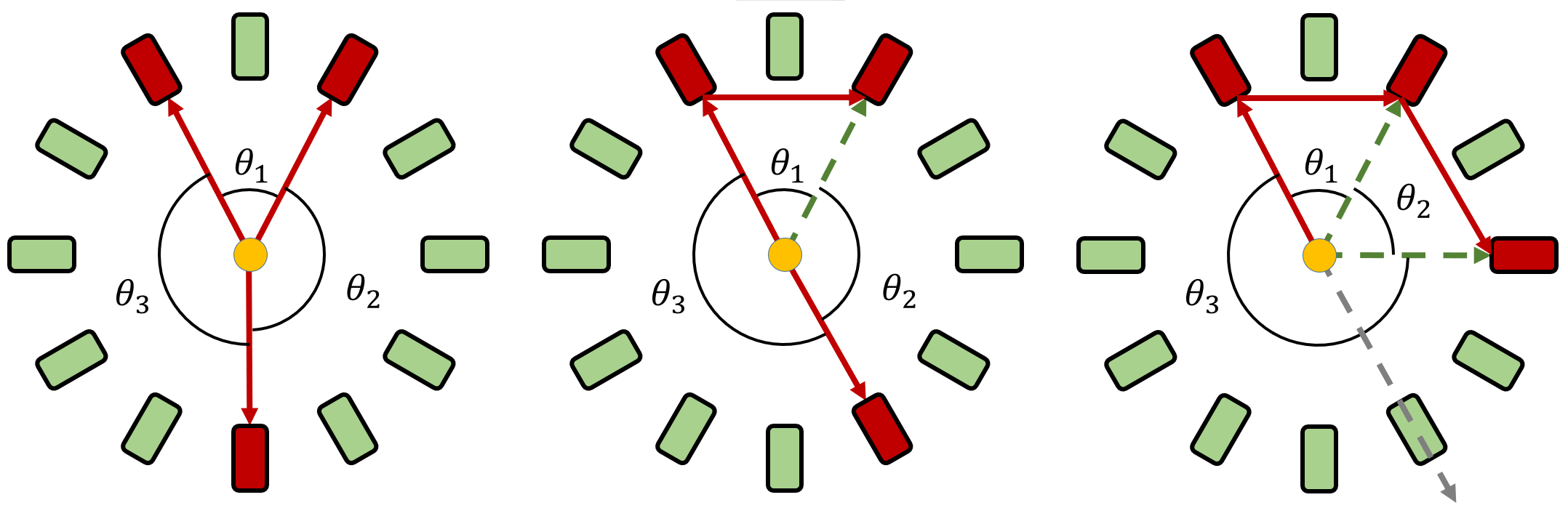}
\caption{\label{fig:Angles_types} Scheme of different kind of events expected in the case of three registered interactions:
positronium or direct electron-positron annihilation into 3 photons (left), into two back-to-back photons with additional registration of secondary scattering of one of them (middle), and into 2 photons when one of them is not registered and the other undergoes double scattering
(right).
For clarity only few scintillators in one detection layer are shown. 
}
\end{figure}

\begin{figure}[b!]
\centering
\includegraphics[width=0.4\textwidth]{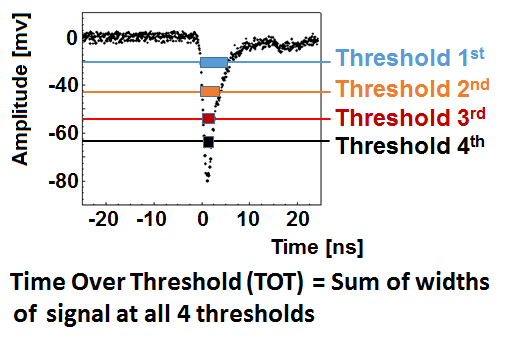}
\includegraphics[width=0.44\textwidth]{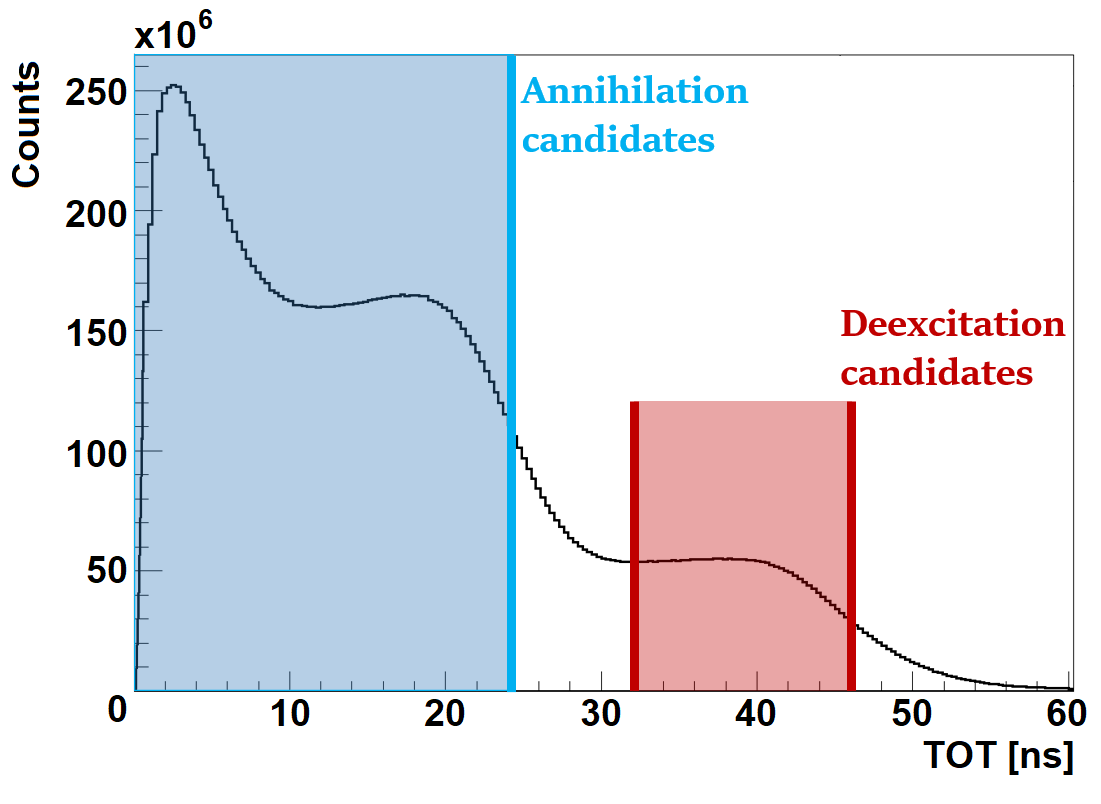}
\caption{\label{fig:TOT} (Left panel) Pictorial definition of the Time-Over-Threshold (TOT) of the signal measured in the J-PET detector. TOT is a sum of signal widths in the time domain, over four preset voltage levels.
(Right panel) Distribution of the TOT values measured by means of the J-PET detector. Categorization of the hit is based on the TOT value. In the analysis, the annihilation candidate is defined when TOT value is less than 24 ns. De-excitation candidates corresponds to the TOT greater than 32 ns and smaller than 46 ns.} 
\end{figure}

Here we concentrated on the exclusive measurement of three photons from the o-Ps $\to$ 3$\gamma$ decay, case III in Table~\ref{Tab:TypesOfEvents}. Such exclusive measurements enable one to reconstruct the full event kinematics, making all photons' four-momentum vectors available for the physics analysis,
suppressing the physical and instrumental background. 
Measurement of the de-excitation photon allows one to determine the lifetime of the positron in the target, and hence to disentangle between the o-Ps decays from the direct annihilation events (e$^+$e$^-$ $\to$ 3$\gamma$), and to suppress the background originating from p-Ps $\to$ 2$\gamma$ decays followed by the secondary scattering of photons in the detector. The latter background is illustrated in the middle and right panels of Fig.~\ref{fig:Angles_types}. De-excitation and annihilation photons are identified based on the measured time-over-threshold (TOT) values~\cite{Sushil_EJNMMI} that are correlated with the energy deposition in the scintillators.

\begin{figure*}[t!]
\centering
\includegraphics[width=0.44\textwidth]{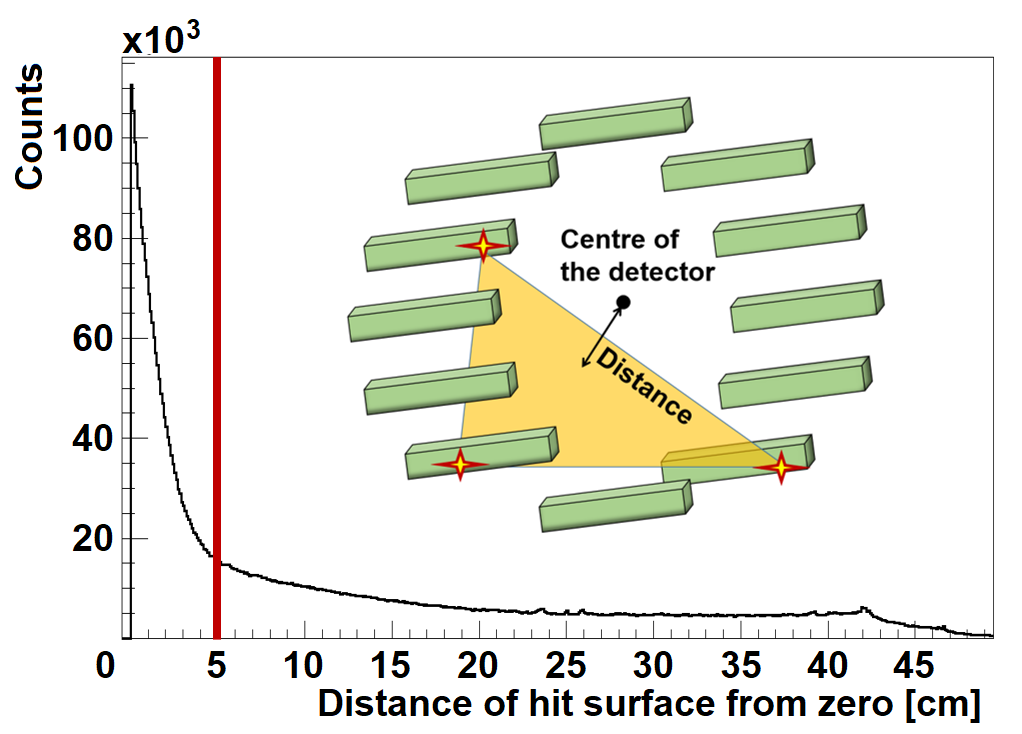}
\includegraphics[width=0.45\textwidth]{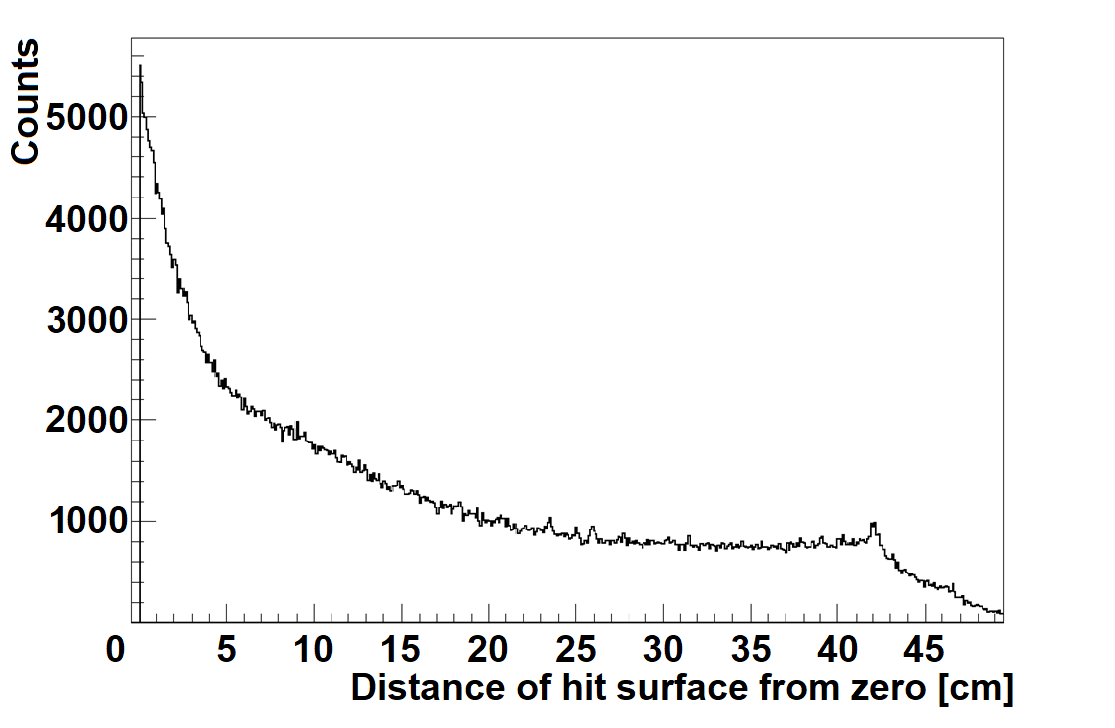}\\
\includegraphics[width=0.44\textwidth]{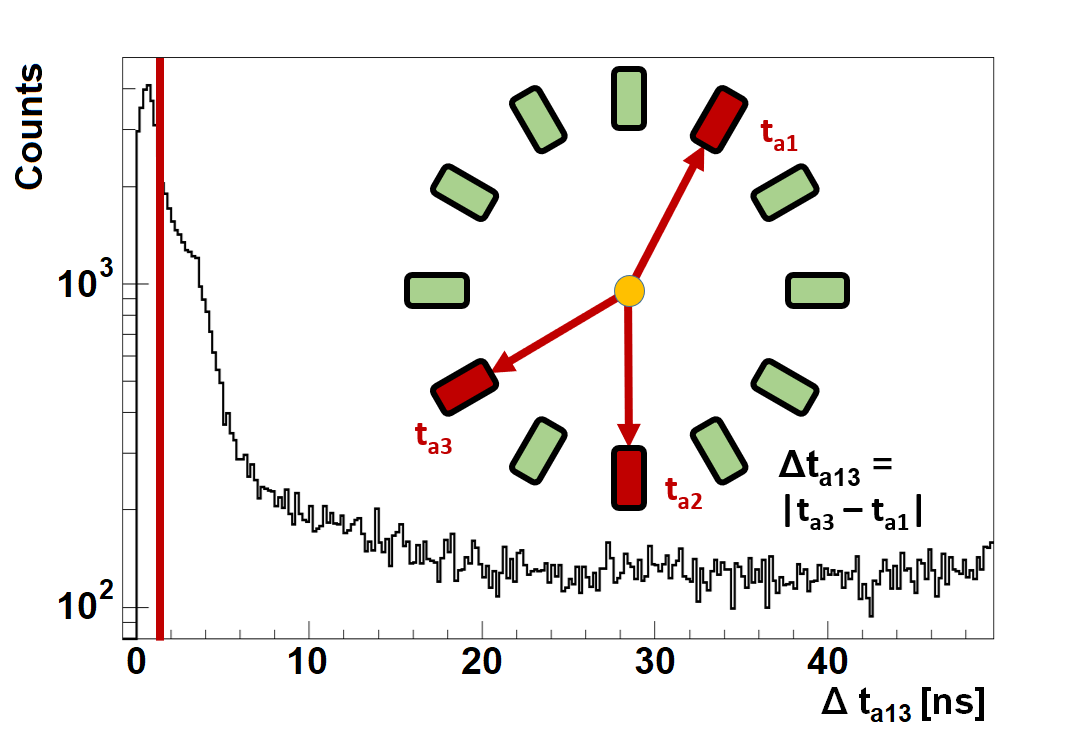}
\includegraphics[width=0.44\textwidth]{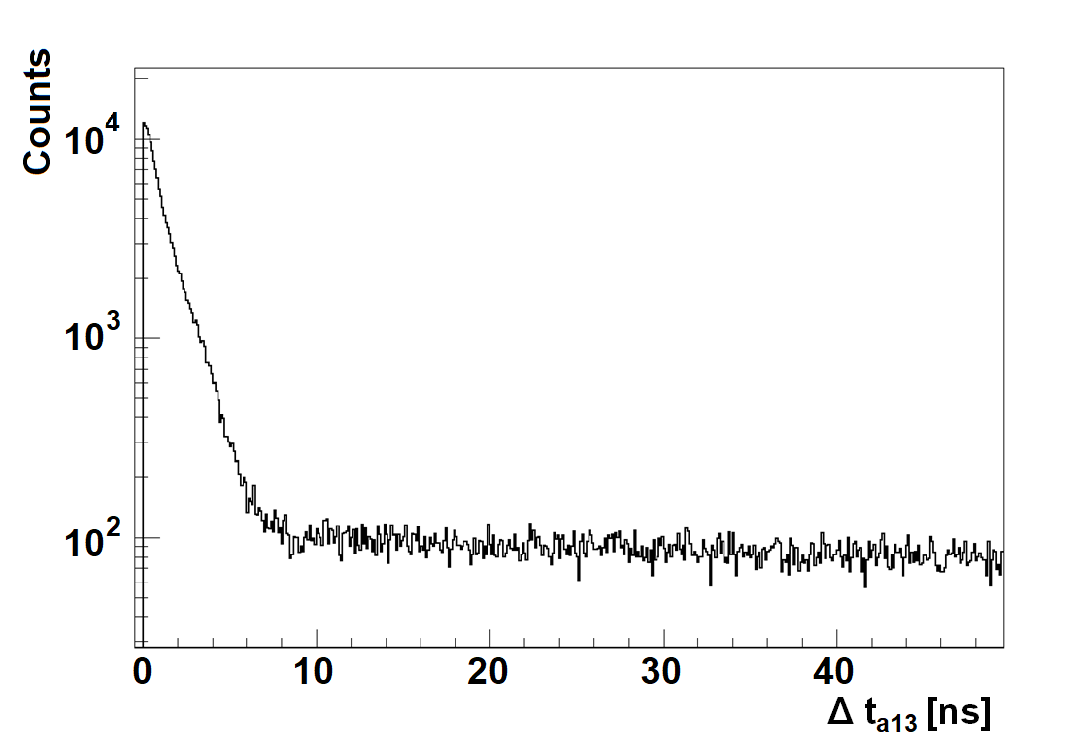}
\caption{\label{fig:DistAnniTDiff} Distributions of the distance of the decay plane from the centre of the detector for the experimental data (upper panel, left) and simulations (upper panel right), and the time difference $\Delta t_{a13}$ between annihilation hits for the experimental data (lower panel, left) and the simulations (lower panel, right).
Red line indicates the condition which is used for suppressing the background.
}
\end{figure*}

Fig.~\ref{fig:TOT} shows a pictorial definition of the TOT values measured with the J-PET front-end electronics~\cite{FPGA2} 
(upper panel)
and an example TOT spectrum with superimposed regions used for the identification of the de-excitation and annihilation photons (lower panel). 
At each threshold (80 mV, 160 mV, 240 mV, 320 mV)
a  TOT$_\text{thr}$ value is obtained, and the
TOT value is calculated as a sum of these TOT$_\text{thr}$. 
Using four threshold values for estimation of the TOT allows for a better estimation of the signal charge. It was shown that the TOT value as in Fig.~\ref{fig:TOT} is a good estimator of the deposited energy~\cite{Sushil_EJNMMI}.
The first level of reduction of background from rescattering (middle and right panel of Fig.~\ref{fig:Angles_types}), from accidental coincidences and from cosmic rays, 
was applied as a~condition
on the distance between the decay plane and the annihilation position, as well as the time difference between the interaction of the annihilation photons' candidates ($\Delta t_{a13}$). 
Due to  momentum conservation, momentum vectors of the photons from the o-Ps $\to$ 3$\gamma$ decay form a plane, 
subsequently referred to as the decay plane. 
In the ideal case the decay plane constructed from the measured interaction points should comprise the annihilation point. Therefore, for the o-Ps $\to$ 3$\gamma$ events, the distance between the decay plane and the centre of the annihilation chamber should be close to zero, while this distance for the background events may spread even up to the radius of the detector, as shown in the upper panel of Fig.~\ref{fig:DistAnniTDiff}. Similarly, the reconstructed value of $\Delta t_{a13}$ should be close to zero in the case of the true o-Ps $\to$ 3$\gamma$  process, where  $\Delta t_{a13} = \left| t_{a3} - t_{a1} \right|$ with $t_{a3}$ - the largest time of registration, and $t_{a1}$ - the smallest time,  out of the three determined emission times of the annihilation photons candidates. In the case of the double scattering (right panel of Fig.~\ref{fig:Angles_types}) $\Delta t_{a13}$ may reach a value of about 6~ns and for accidental coincidences it forms a continuous flat distribution as can be seen in the lower panel of Fig.~\ref{fig:DistAnniTDiff}.

\begin{figure}[h!]
\centering
\includegraphics[width=0.43\textwidth]{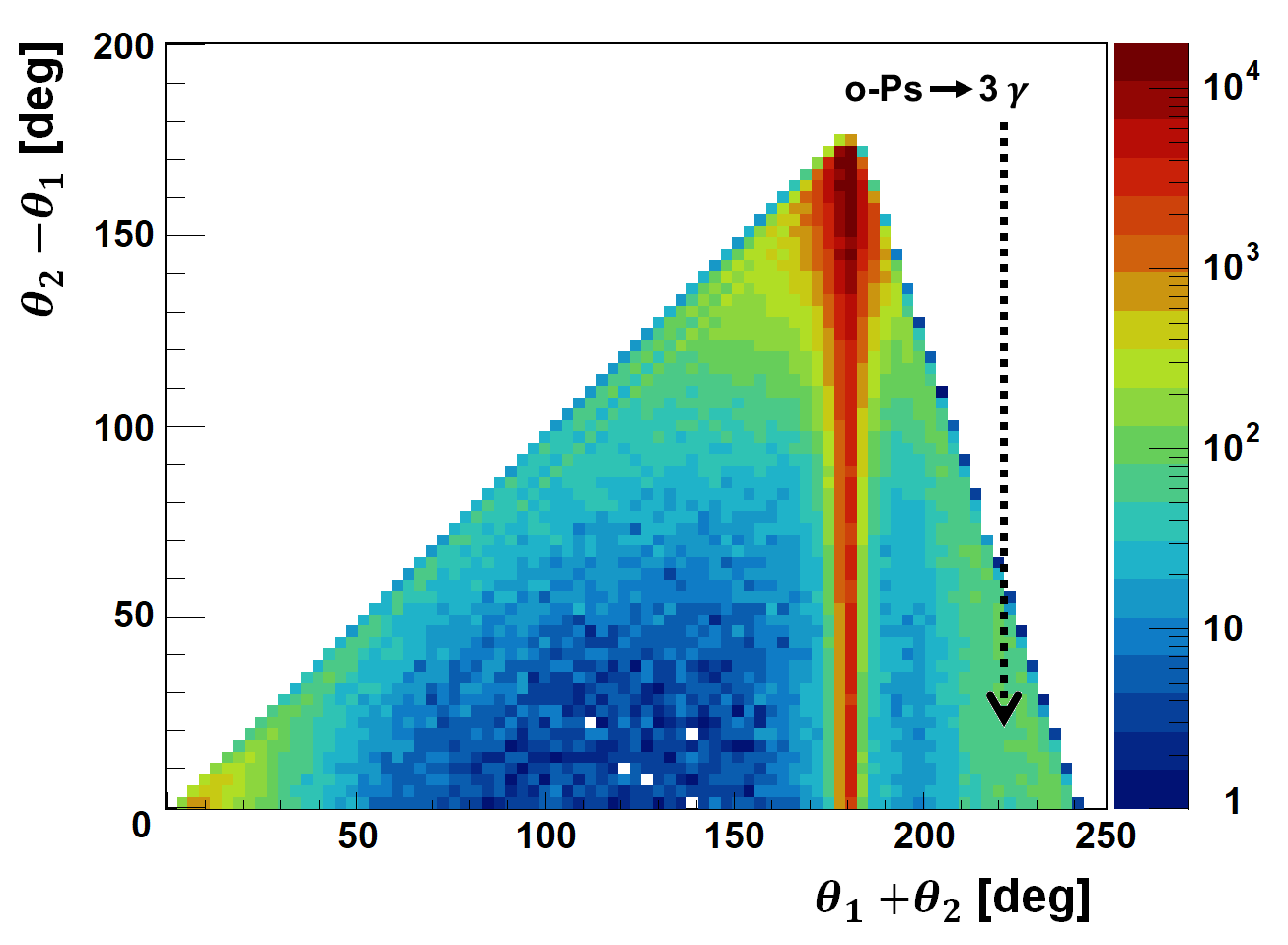}
\includegraphics[width=0.52\textwidth]{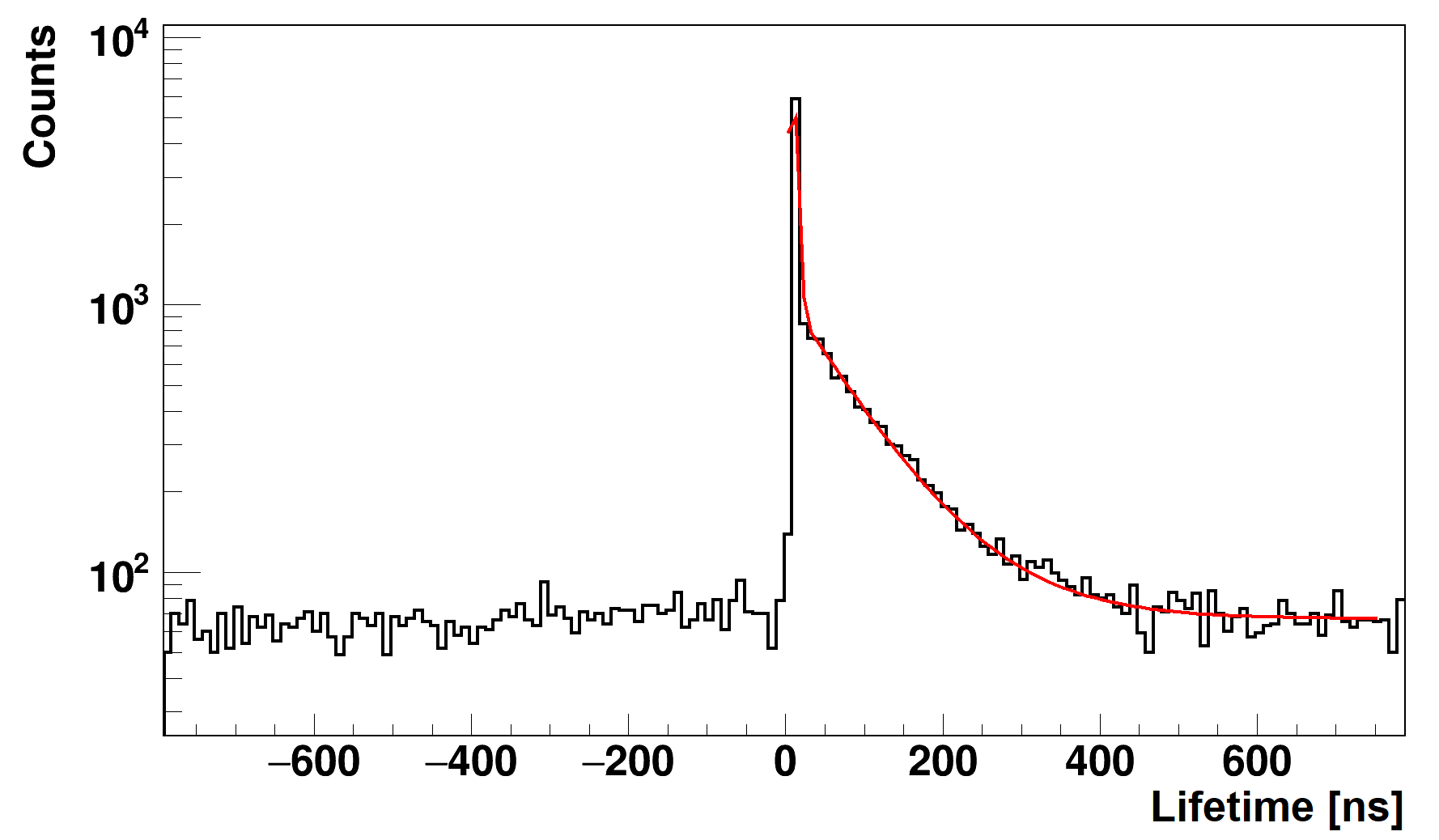}\\
\caption{ \label{Angles_LFComp} (Left panel) Distribution of the difference versus sum of the two smallest relative angles between reconstructed momentum vectors of the three annihilation photon candidates for the experimental data. The most pronounced vertical band at 180 degrees corresponds to events with annihilations into two back-to-back photons with additional scattering of one of them~(middle panel of Fig.\ref{fig:Angles_types}). Events on the left side of this band correspond predominantly to the double scattering (right panel of Fig.\ref{fig:Angles_types}) and the signal from o-Ps $\to$ 3$\gamma$ process is visible as an enhancement indicated by an arrow. 
(Right panel) Positron lifetime distribution determined for events that passed selection criteria for experimental data. The lifetime was calculated as the difference between the mean time of the emission of annihilation photons and the time of the emission of the de-excitation photon. Red line indicates fit to the data.}
\end{figure}

Further, at the next stage of the background reduction and 
identification of the o-Ps $\to$ 3$\gamma$ events, the angular correlation presented in Fig.~\ref{Angles_LFComp} is used. The most pronounced single-scattered background events \linebreak (middle panel of Fig.~\ref{fig:Angles_types}) form a band at 180 degrees, while the double-scattered background events (right \linebreak panel of Fig.~\ref{fig:Angles_types}) are concentrated predominantly at the left side of this band. For the true o-Ps $\to$ 3$\gamma$ decays, due to the momentum conservation, the sum of the two smallest relative angles defined in Fig.~\ref{fig:Angles_types} is larger than 180 degrees and the distribution has the maximum for the symmetric configuration with the sum of relative angles equal to 240 degrees~\cite{X5,Kaminska:2016fsn}. In Fig.~\ref{Angles_LFComp} (left panel) which shows experimental data obtained with the J-PET detector, such enhancement is clearly visible as expected. 
In order to suppress the background originating from the  back-to-back photons, it is required that the sum of two smallest angles is larger than 190 degrees.
Further, as the last stage of the selection, a~lifetime histogram is used.

Fig.~\ref{Angles_LFComp} presents our main result,
the lifetime spectrum for 3$\gamma$ events candidates meeting all of the selection criteria described. The signal from o-Ps $\to$ 3$\gamma$ process is seen as a clear exponential decay with the mean lifetime of $90.2 \pm 3.5$ ns consistent with the value of $90.8 \pm 1.2$ ns expected for the mean lifetime of ortho-positronium in the used XAD4 porous polymer~\cite{Jasin}. In addition, the production intensity of the o-Ps component (fraction of the o-Ps 3$\gamma$ decays in the distribution $\rightarrow$ signal from 3$\gamma$) is in agreement between the experiment ($52.8 \pm 1.5 \%$) and the predictions ($53.7 \pm 3.1 \%$). The predictions were made on the basis of simulations that included different fractions of positron-electron decays in XAD4 \cite{Jasin}, detectors acceptance and analysis stream.
This spectrum enables one to reject the background from direct  e$^+$e$^-$ $\to$ 3$\gamma$ annihilation, for which the lifetime is equal to about 0.5~ns~\cite{Jasin}. Also the remaining background from the cosmic radiation and the scatterings in the annihilation chamber concentrates within few nanoseconds around zero (the maximum time needed for cosmic ray to pass through the detector). Fraction of the cosmic rays was estimated based on the separate measurement and it was equal to $1\%$. Taking into account corrected intensity value given above, cosmic rays can explain possible differences between simulations and experiment.

The flat distribution below the o-Ps lifetime spectrum for the negative 
lifetime values corresponds to the background events due to the accidental coincidences \cite{FPGA1}. This feature enables one to estimate
statistically the contribution from the accidental coincidences to the distributions which will be used for the physics studies (such as e.g. Dalitz plot distribution of the o-Ps $\to$ 3$\gamma$ process). For the spectrum shown in Fig.~\ref{Angles_LFComp}, 
based on the simulation we can distinguish different fractions of events: clean events coming from the o-Ps annihilation into 3$\gamma$ (43.01$\%$); events in which we observe scattering of one of three annihilation photons (25.06$\%$); events in which one annihilation hit is mistaken with hit coming from the scattering of the de-excitation photon (1.69$\%$); events in which we register positron annihilation into 2$\gamma$ with one additional scattering of annihilation or de-excitation photon (22.41$\%$); mixed event of different annihilations (7.82$\%$).

\section{Conclusions}
\begin{figure}
\centering
\includegraphics[width=0.48\textwidth]{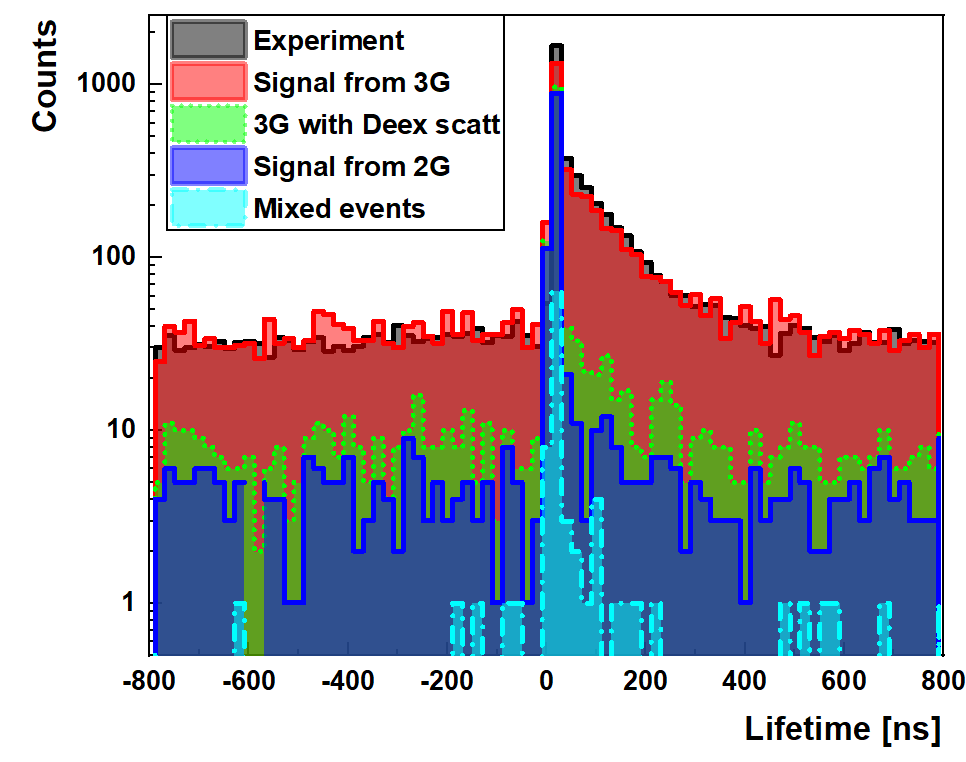}
\caption{ \label{LF_Decomp} Comparison of the positron lifetime spectrum between the experiment (black line) and the simulations. Simulation spectrum was divided into different components with different origins: annihilations of positron-electron into 3 photons (red line), events in which one or more annihilation hit was coming from the scattering of the deexcitation photon (green line), annihilations of positron-electron into 2 photons with scattering of the deexcitation photon (blue line), and the mix of the different decays and more complicated scatterings containing secondary particles (cyan line). Experimental spectrum was normalized to the simulations. Possible discrepancies between the experimental data and the simulations (at the peak around zero) can be explained by the cosmic ray background. From a separate measurement, the effect of cosmic rays was estimated at about 1$\%$. }
\end{figure}

In this article we have shown 
that the J-PET tomography system constructed solely from plastic scintillator detectors is capable of exclusive measurements of the decays of ortho-positronium atoms. The elaborated selection method is based on the measurement of the time, position and energy deposition of photons in the scintillator strips, in contrast to studies conducted so far with the crystal and semiconductor based detection systems where the key selection of events is based just on the measurement of the photons' energies~\cite{X5,X4,X2,Yamazaki:2009hp,Y1,Y2,X1,Bernreuther:1988tt}. 
In addition, based on the Monte Carlo simulations we have estimated various background sources, that can survive selection procedure, which allows to properly correct the lifetime distribution.
The presented results, such as shown in Fig.~\ref{LF_Decomp},
demonstrate that J-PET detector enables one to register o-Ps decays 
into 3 photons and to disentangle these decays from the background processes. 
Proper use of selection criteria may allow the simultaneous 
study of other types of positronium decays. 
With the detector performance established here,
the J-PET detector provides a new tool for fundamental physics studies of positronium decays \cite{Moskal:2016moj}, for positron annihilation lifetime spectroscopy, and for entanglement studies \cite{Hiesmayr_ScientiReport2017,Hiesmayr_ScientificReport2019}. Information about the positronium lifetime in biological materials could also open a new perspective for application of positronium as a diagnostic biomarker for medicine~\cite{nrp:2019,Moskal:2018wfc,Mos_PETClinics,EJNMMI_arXiv:1911.06841}.


\section*{Acknowledgements}
The authors acknowledge technical and administrative support of A. Heczko, M. Kajetanowicz and W. Migda\l{}. This work was supported by The Polish National Center for Research
and Development through grant INNOTECH-K1/IN1/64/159174/NCBR/12, the
Foundation for Polish Science through the MPD and TEAM POIR.04.04.00-00-4204/17 programmes, the National Science Centre of Poland through grants no.\
2016/21/B/ST2/01222,\linebreak[3] 2017/25/N/NZ1/00861,
the Ministry for Science and Higher Education through grants no. 6673/\-IA/\-SP/\-2016,
7150/E-338/SPUB/2017/1,
N17/\-MNW/\-000001, N17/MNW/000005, N17/MNS/000017, N17/MNS/000023, and N17/\-MNS/\-000036
the Au\-strian Science Fund FWF-P26783,
the EU Horizon 2020 research and innovation programme, STRONG-2020 project, under grant agreement No 824093,
and
the SciMat Priority Research Area  budget under the program {\it Excellence Initiative - Research University} at the Jagiellonian University.


\bibliographystyle{plain}

\begin{thebibliography}{99}

\bibitem{Moskal:2014sra}
  P.~Moskal {\it et al.},
  Nucl.\ Instrum.\ Meth.\ A {\bf 764} (2014) 317.
%
\bibitem{Moskal:2015nim}
  P.~Moskal {\it et al.},
  Nucl.\ Instrum.\ Meth.\ A {\bf 775} (2015) 54.
%

\bibitem{X5}
%
A. P. Mills and S. Berko, 
Phys. Rev. Lett. {\bf 18} (1967) 420.
%
\bibitem{X4}
%
J. Yang {\it et al.}, 
Phys. Rev. A {\bf 54} (1996) 1952.
%
\bibitem{X2}
%
P. A. Vetter and S. J. Freedman, 
Phys. Rev. A {\bf  66} (2002) 052505.
%

\bibitem{Yamazaki:2009hp}
T.~Yamazaki, T.~Namba, S.~Asai and T.~Kobayashi,
Phys.\ Rev.\ Lett.\  {\bf 104} (2010) 083401;
Erratum: [Phys.\ Rev.\ Lett.\  {\bf 120} (2018) 239902].

\bibitem{Y1} 
M. Skalsey and J. Van House, 
Phys. Rev. Lett. {\bf 67} (1991) 1993.

\bibitem{Y2} 
B.K. Arbic et al., Phys. Rev. {\bf A 37} (1988) 3189.

\bibitem{X1}
P. A. Vetter and S. J. Freedman, 
Phys. Rev. Lett. {\bf 91} (2003) 263401.

\bibitem{Bernreuther:1988tt}
  W.~Bernreuther, U.~Low, J.~P.~Ma and O.~Nachtmann,
  Z.\ Phys.\ C {\bf 41} (1988) 143.

\bibitem{Moskal:2018wfc}
P.~Moskal {\it et al.},
Phys. Med. Biol. \textbf{64} 5 (2019) 055017.

\bibitem{Mos_PETClinics}
P. Moskal and E.Ł. Stępień,
PET Clinics {\bf 15} (2020) 439.

\bibitem{VanMosKar_EJNMMI}
S. Vandenberghe, P. Moskal, J. Karp, 
EJNMMI Phys. 7 (2020) 35.

\bibitem{SloPanGer_SNM}
P. J. Slomka, T. Pan and G. Germano, 
 Semin. Nucl. Med., vol. \textbf{46} (2016) 5.

\bibitem{Badawi_JNM}
R.D. Badawi \textit{et al.},
J. Nucl. Med. \textbf{60}(3) (2019) 299.

\bibitem{Karp_JNM}
J.S. Karp \textit{et al.}, 
J. Nucl. Med. \textbf{61}(1) (2020) 136.

\bibitem{Moskal:2016a} 
P. Moskal {\it et al.}, 
Phys. Med. Biol. {\bf 61} (2016) 2025.
%
\bibitem{Niedzwiecki:2017app}
Sz. Niedźwiecki {\it et al.}, Acta Phys. Polon. B {\bf 48} (2017) 1567.

\bibitem{Kowalski:2018pmb}
P. Kowalski {\it et al.}, Phys. Med. Biol. {\bf 63} (2018) 165008.

\bibitem{Moskal:2016moj}
  P.~Moskal {\it et al.},
  Acta Phys.\ Polon.\ B {\bf 47} (2016) 509.

\bibitem{Kaminska:2016fsn}
D.~Kamińska
{\it et al.}, 
Eur. Phys. J. C \textbf{76} 
(2016) 445.

\bibitem{Moskal:2018pus}
P.~Moskal
{\it et al.}, 
Eur. Phys. J. C \textbf{78} 
(2018) 970.


\bibitem{Cassidy:2018tgq} 
D.~B.~Cassidy,
Eur.\ Phys.\ J.\ D {\bf 72} (2018) 53.

\bibitem{Bass:2019ibo}
S.~D.~Bass,
Acta Phys. Polon. B \textbf{50} (2019) 1319.

\bibitem{Adkins:2002fg}
G.~S.~Adkins, R.~N.~Fell and J.~Sapirstein,
  Ann.\ Phys.\ {\bf 295} (2002) 136.

\bibitem{Nikodem:2019} N. Krawczyk \textit{et al.}, Hyperfine Interact 240 (2019) 81.

\bibitem{Krzemien_Elena_Kacprzak_2020_Acta}
W. Krzemień, E. Pérez del Río, K. Kacprzak
Acta Phys. Pol. B \textbf{51} (2020) 165.

\bibitem{Vigo:2018xzc}
  C.~Vigo, L.~Gerchow, L.~Liszkay, A.~Rubbia and P.~Crivelli,
  Phys.\ Rev.\ D {\bf 97} (2018) 092008.
  
\bibitem{Vigo:2020}
  C.~Vigo \textit{et al.}, Phys.\ Rev.\ Lett. {\bf 124} (2020) 101803.

\bibitem{nrp:2019}
P. Moskal, B. Jasińska, E.Ł. Stępień and S.D. Bass,
Nat. Rev. Phys. \textbf{1} (2019) 527.

\bibitem{bura:2020app}
Z. Bura \textit{et al.}, Acta Phys. Polon. B \textbf{51} (2020) 377.

\bibitem{jasin:2017appa}
B. Jasińska \textit{et al.}, Acta Phys. Polon. A 132 (2017) 1556.

\bibitem{EJNMMI_arXiv:1911.06841}
P.~Moskal {\it et al.},
EJNMMI Phys. {\bf 7} (2020) 44.

\bibitem{Hiesmayr_ScientiReport2017} 
B.C. Hiesmayr, P. Moskal, Sci. Rep. {\bf 7} (2017) 15349.

\bibitem{Hiesmayr_ScientificReport2019} 
B.C. Hiesmayr, P. Moskal, Sci. Rep. {\bf 9} (2019) 8166. 

\bibitem{Mat1}
J.N. Sun, Y.F. Hu, W.E. Frieze, D.W. Gidley, Radiation Physics and Chemistry \textbf{68} (2003) 345.

\bibitem{Mat2}
Y. Kobayashi, K. Ito, T. Oka, K. Hirata, Radiation Physics and Chemistry \textbf{76} (2007) 224.

\bibitem{EJ230} https://eljentechnology.com, Plastic scintillator EJ-230.
\bibitem{R9800} http://hamamatsu.com, Photomultiplier tube R9800s

\bibitem{Neha_reco} N.G. Sharma \textit{et al.}, IEEE Transactions on Radiation and Plasma Medical Sciences, Vol. 4, No. 5 (2020) 528.

\bibitem{Lech_reco} L. Raczyński \textit{et al.}, Phys. Med. Biol. 62 (2017) 5076.

\bibitem{FPGA2} M. Pałka \textit{et al.}, JINST \textbf{12} (2017) P08001.
\bibitem{FPGA1} G. Korcyl \textit{et al.}, IEEE Transactions On Medical Imaging \textbf{37} (2018) 2526.

\bibitem{Framework} W. Krzemień \textit{et al.}, Acta Phys. Polon. B \textbf{47} (2016) 561.

\bibitem{Framework2} W. Krzemień \textit{et al.}, 
SoftwareX 11 (2020) 100487.
%

\bibitem{APEX}
C. Bartram, R. Henning, D. Primosch, Nuclear Inst. and Methods in Physics Research A \textbf{966} (2020) 163856.

\bibitem{sigma}
{\tt https://www.sigmaaldrich.com}, CAS Number 37380-42-0.

\bibitem{Jasin} B. Jasińska \textit{et al.},  Acta Phys. Polon. B \textbf{47} (2016) 453.

\bibitem{Gorgol_APB}
M. Gorgol \textit{et al.}, Acta Phys. Pol. B \textbf{51} (2020) 293

\bibitem{Step_PCCP}
P.V. Stepanov \textit{et al.},
Phys. Chem. Chem. Phys. \textbf{22} (2020) 5123.

\bibitem{Pick_off} R.L. Garwin, Phys. Rev. \textbf{91} (1953) 1571.

\bibitem{PALS_Review}
Y.C. Jean, P.E. Mallon and D.M. Schrader, 
{\it Principles and Applications of Positron and Positronium Chemistry} (World Scientific, 2003) 1.

\bibitem{Sushil_EJNMMI}
S. Sharma \textit{et al.},
EJNMMI Physics (2020) 7.


\end{thebibliography}

\end{document}